%% file: bullsrsl-sample-en.tex
\documentclass[final,english]{bullsrsl}[2022/06/15]
\usepackage{ifthen}
\usepackage{mathptmx}
\usepackage{hyperref}
\usepackage{lineno}
\usepackage{geometry}
\usepackage{xspace}

\usepackage[latin1]{inputenc}
\usepackage[T1]{fontenc}
\usepackage{natbib} 
\usepackage{graphicx}
\newcommand{\fermi}{{\em Fermi}\xspace}
\newcommand{\swift}{{\em Swift}\xspace}
\newcommand{\sw}[1]{\texttt{#1}}

\newcommand{\apjl}{ApL}

\newcommand{\aap}{A\&A}
\newcommand{\apjs}{ApJS}

\begin{document}
\title{Magnetar central engine powering the energetic GRB 210610B ?}
\author[affil={1,2}, corresponding]{Amit}{Kumar Ror}
\author[affil={1}]{Shashi}{B. Pandey}
\author[affil={1,3}]{Rahul}{Gupta}
\author[affil={1,3}]{Amar}{Aryan}
\affiliation[1]{Aryabhatta Research Institute of Observational Sciences (ARIES), Nainital-263002, India}
\affiliation[2]{Department of Applied Physics, Mahatma Jyotiba Phule Rohilkhand University, Bareilly-243006, India}
\affiliation[3]{Department of Physics, Deen Dayal Upadhyaya Gorakhpur University, Gorakhpur-273009, India}
\correspondance{amitror@aries.res.in}
\maketitle

\vspace{-0.5cm}
\begin{abstract}
The bright GRB 210610B was discovered simultaneously by \fermi and \swift missions at redshift 1.13. We utilized broadband \fermi-GBM observations to perform a detailed prompt emission spectral analysis and to understand the radiation physics of the burst. Our analysis displayed that the low energy spectral index ($\alpha_{\rm pt}$) exceeds boundaries expected from the typical synchrotron emission spectrum (-1.5,-0.67), suggesting additional emission signature. We added an additional thermal model with the typical \sw{Band} or \sw{CPL} function and found that \sw{CPL} + BB function is better fitting to the data, suggesting a hybrid jet composition for the burst. Further, we found that the beaming corrected energy (E$_{\rm \gamma, \theta_{j}}$ = 1.06 $\times$ 10$^{51}$ erg) of the burst is less than the total energy budget of the magnetar. Additionally, the X-ray afterglow light curve of this burst exhibits achromatic plateaus, adding another layer of complexity to the explosion's behavior. Interestingly, we noted that the X-ray energy release during the plateau phase (E$_{\rm X,iso}$ = 1.94 $\times$ 10$^{51}$ erg) is also less than the total energy budget of the magnetar. Our results indicate the possibility that a magnetar could be the central engine for this burst.
\end{abstract}

\vspace{-0.5 cm}
\section{Introduction}
Gamma-ray bursts (GRBs) are the most energetic explosions in the universe and emit electromagnetic radiation in two phases. The initial "prompt emission phase" is a complex function of time, energy, and polarization, spanning a broad range of frequencies from radio to gamma rays up to TeV energies \citep{2019Natur.575..455M}, that poses significant challenges for relating it to known physical emission processes. The synchrotron emission mechanism can explain the observed spectral features in some GRBs. However, some bursts exhibit a spectral component that appears to be inconsistent with synchrotron emission, such as the low energy spectral index $\alpha_{\rm pt}$ of the \sw{Band} function not always remaining within the limits (-1.5,-0.67) known as the synchrotron line of death \citep{1998ApJ...506L..23P}. GRBs deviating from these limits can be explained by adding a thermal component to the spectrum, indicating the hybrid jet composition for these bursts \citep{zhang, 2015AdAst2015E..22P}. Following this, the "afterglow emission phase" of a GRB is generated through the synchrotron radiation from the electrons that are accelerated in the external shock formed by the interaction of the GRB ejecta with the surrounding medium. Theoretical models based on this mechanism have been able to reproduce the observed afterglow light curves and spectral properties with remarkable accuracy. However, some observed deviations from the expected afterglow emission, such as achromatic plateaus or bumps, continue to challenge the \sw{synchrotron} model. In recent years, a handful of GRBs have been identified that exhibit a plateau phase in their afterglow light curves. The plateau phase is believed to be a result of energy injection into the external shock, which maintains the shock's constant energy over a longer timescale. The injection can be achieved through different mechanisms; the most plausible scenario is the magnetar central engine, where the energy injection from a magnetar can cause the external plateau in the afterglow light curves \citep{2018ApJ...869..155S}. The cosmological constants chosen for this article are Hubble parameter H$_0=71 $ $\rm km$ $\rm s^{-1} Mpc^{-1}$, density parameters $\Omega_{\Lambda}=0.73,$ and $\Omega_{\rm m}=0.27$.

\vspace{-0.5 cm}
\section{Data Analysis and Results}
For the present analysis, we have used the publicly available data of GRB 210610B, which was detected by the \fermi \citealt{2021GCN.30199....1M} on 2021-06-10 19:51:05.05 UT along with \swift and several other space- and ground-based telescopes during the prompt and afterglow phase.\\

\vspace{-0.6 cm}
\textbf{{\large Prompt emission:}} The Gamma-Ray Burst Monitor (GBM, \citealt{Meegan}) onboard \fermi is specifically designed to detect the prompt emissions of GRBs with excellent temporal and spectral resolution. For analyzing the GBM data, we utilized the Multi-Mission Maximum Likelihood (\sw{3ML}, \citealt{2015arXiv150708343V}) framework. We downloaded the GBM data from the Fermi Science Support Center (FSSC) and utilized time-tagged event files from 3 NaI detectors and 1 BGO detector with the minimum deviation from the direction of the burst. The time-integrated and time-resolved spectra were then extracted using the \sw{gtburst} package. To create the time-resolved spectrum, time slicing was performed by utilizing the Bayesian block binning method with a false alarm probability of 0.01. The extracted spectrum was loaded into \sw{3ML} utilizing the GBM plugin, and we employed various inbuilt empirical models, such as \sw{Band} and \sw{Cuttoff powerlaw} (CPL), along with physical models like blackbody (BB) and physical synchrotron \citep{2020NatAs...4..174B}, and their combinations for spectral fitting. We adopted the Bayesian method to fit the model to the data and evaluated the goodness-of-fit using the Deviance Information Criterion (DIC) statistical test \citep{Spiegelhalter2002}. The model with the lowest DIC value was considered the best fit. For further details about the data analysis and model comparison, kindly refer to \cite{2023ApJ...942...34R}. \\

\begin{figure*}
    \centering
    \includegraphics[scale=0.24]{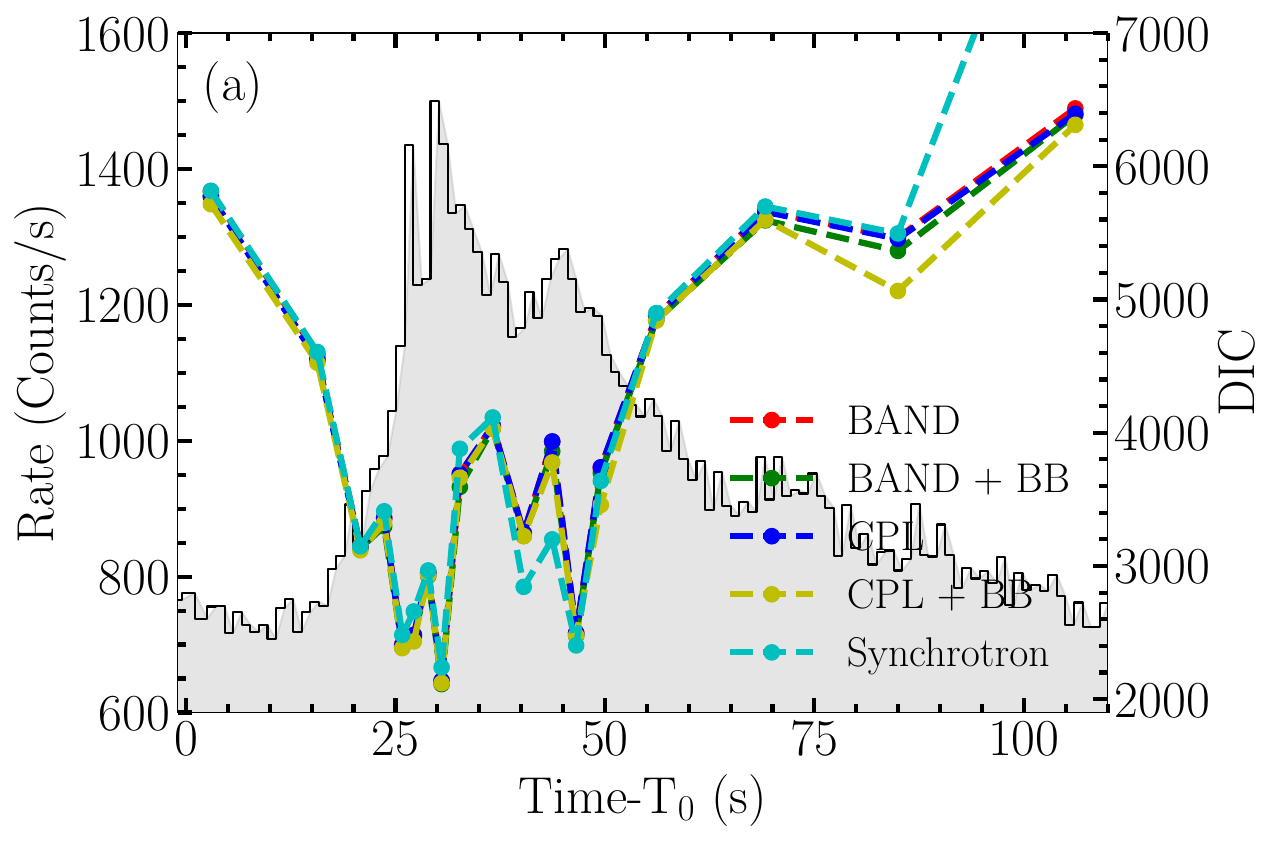}
    \includegraphics[scale=0.24]{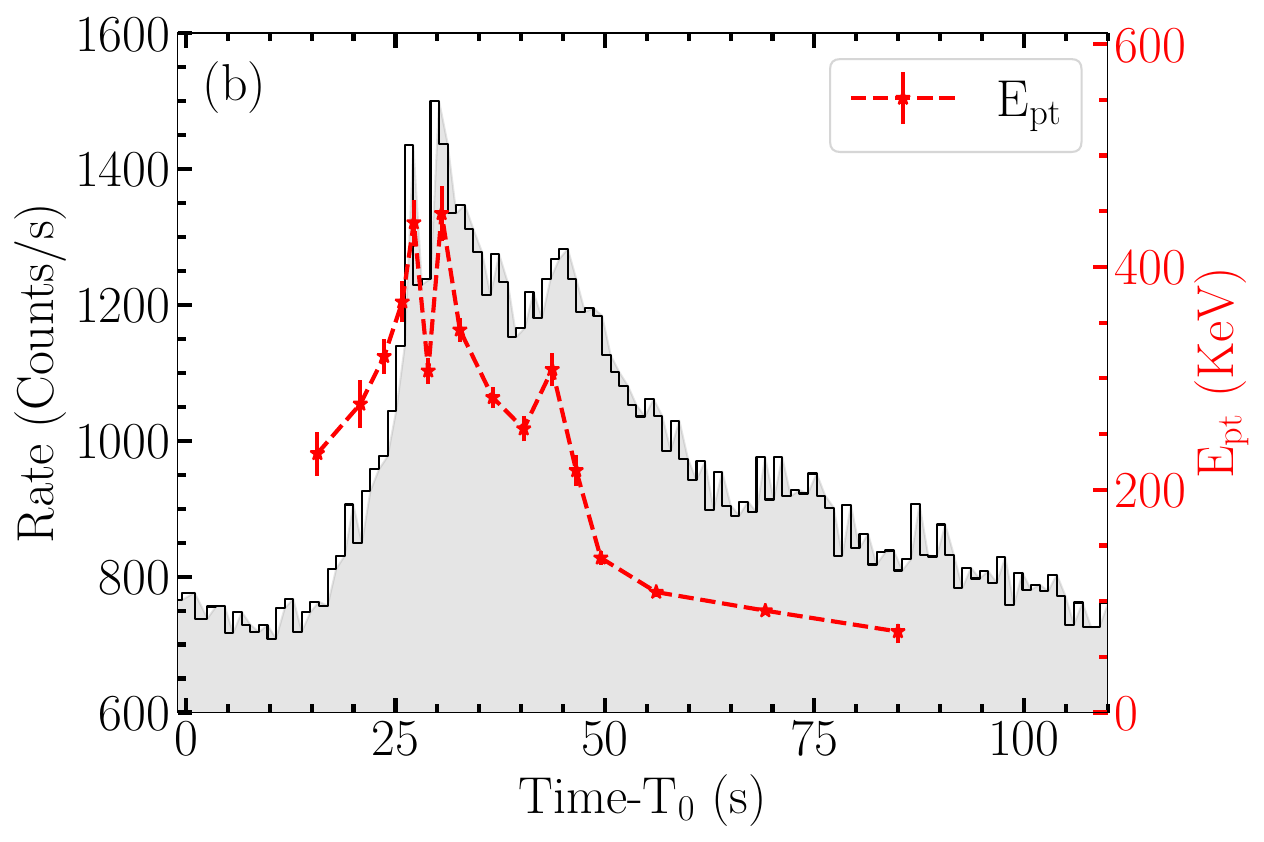}
    \includegraphics[scale=0.24]{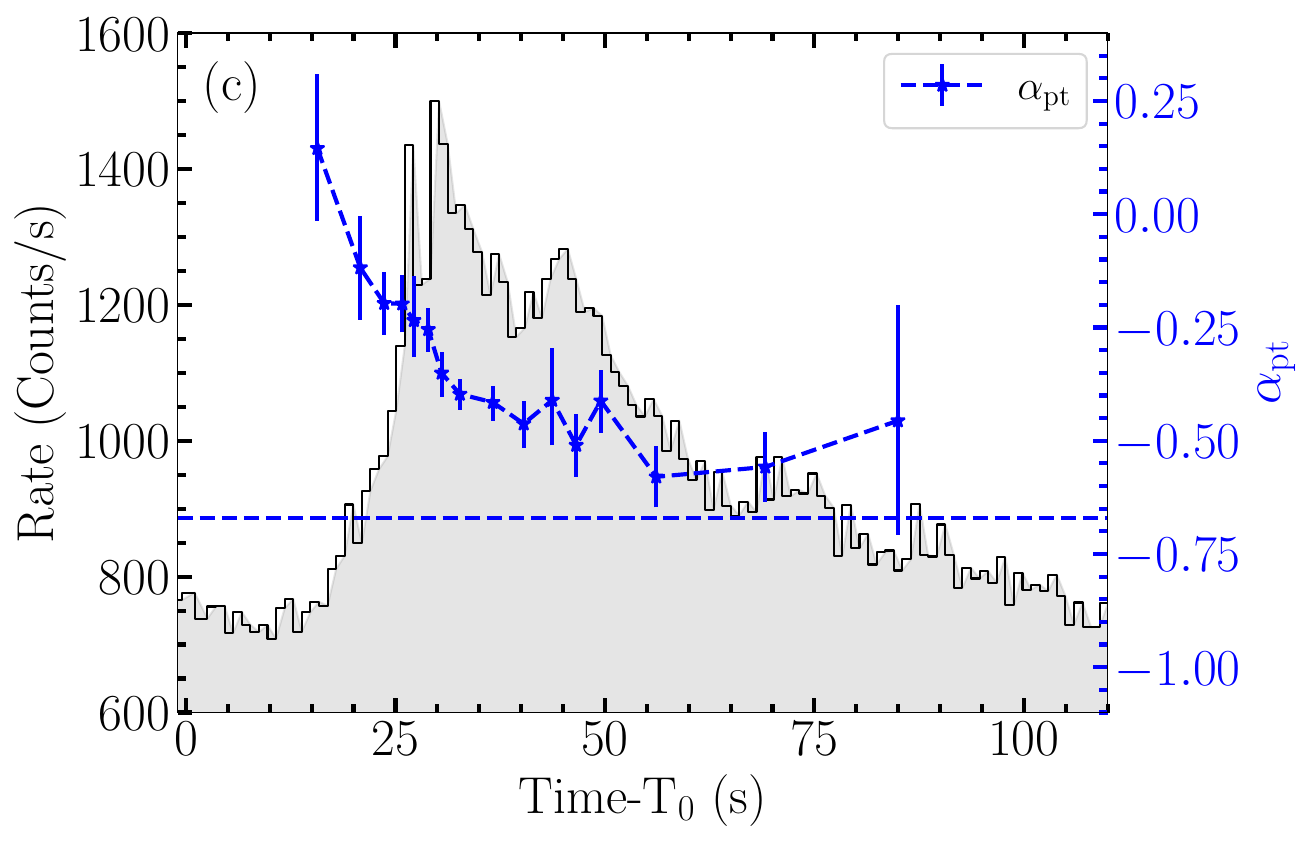}
    \includegraphics[scale=0.24]{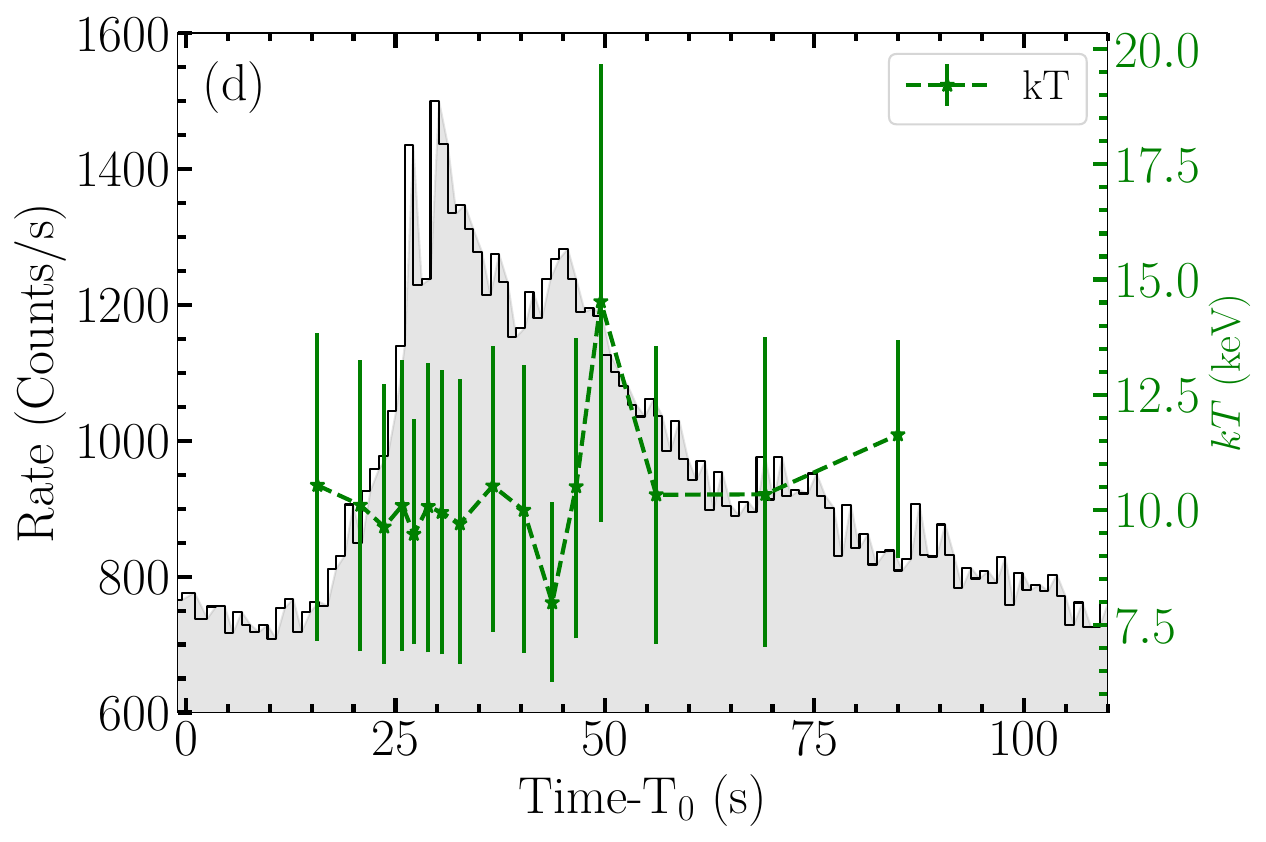}
    \includegraphics[scale=0.24]{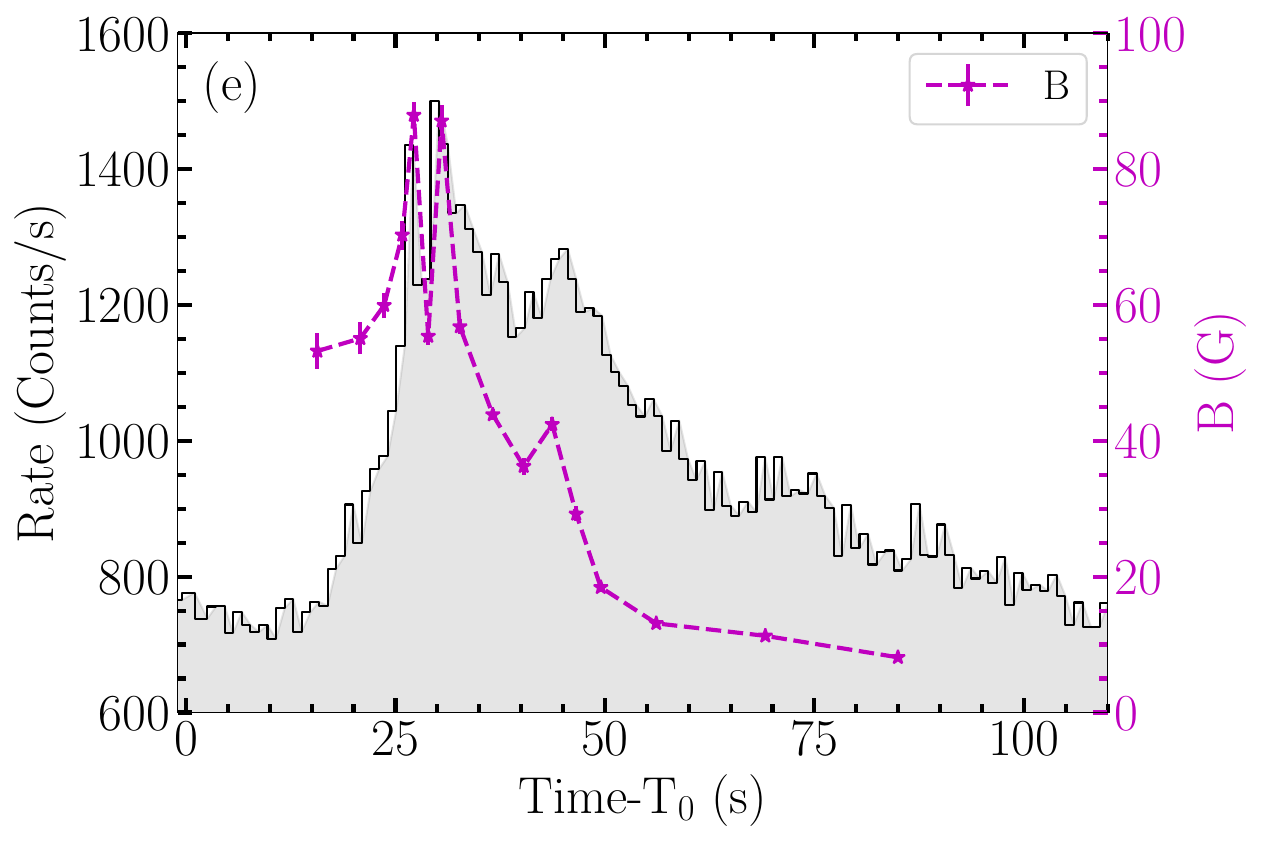}
    \includegraphics[scale=0.24]{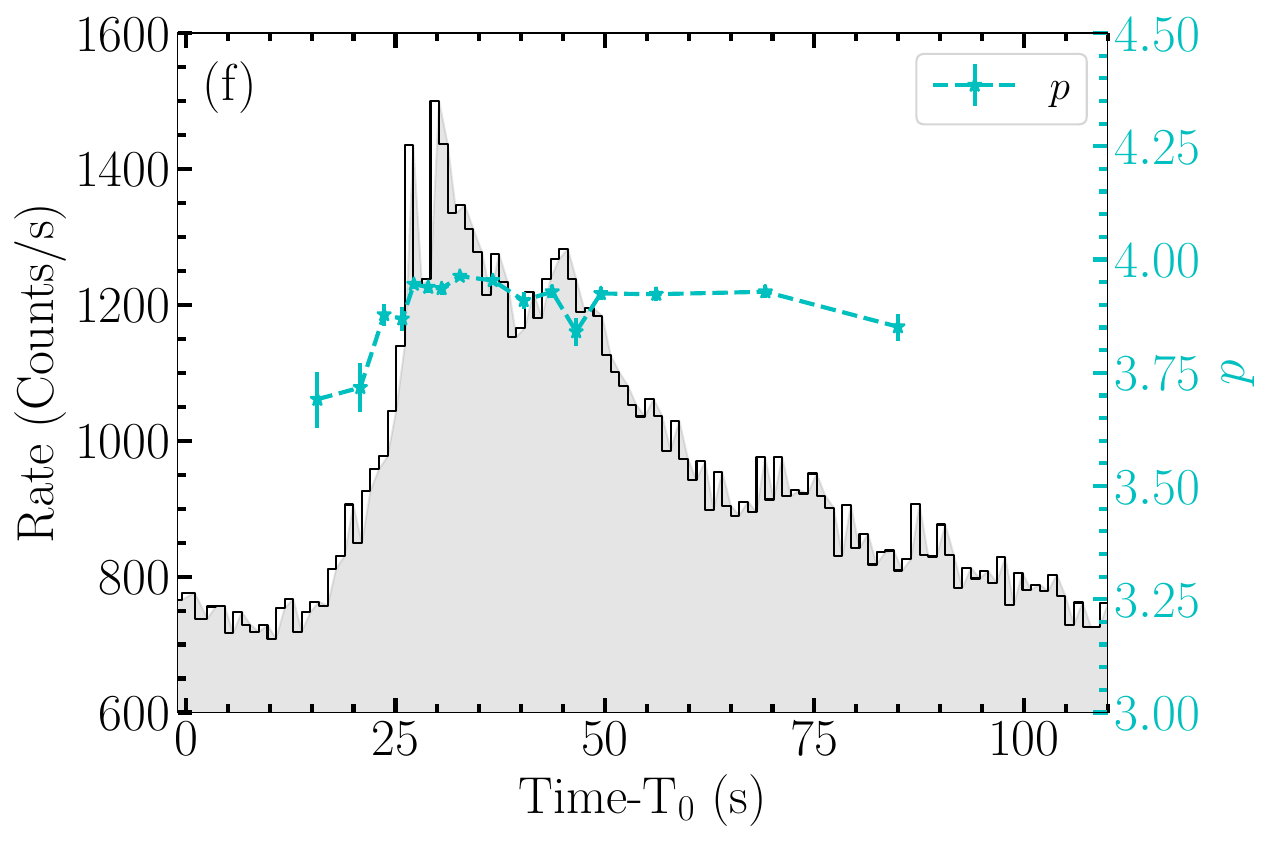}
    \bigskip
    \begin{minipage}{12cm}
    \caption{Results of the time-resolved spectral analysis of GRB 210610B. (a) represents the DIC comparison of various models used to fit the time-resolved prompt emission spectrum. The panel (b), (c), and (d) represent the evolution of peak energy E$_{\rm pt}$, low energy spectral index $\alpha_{\rm pt}$, and temperature (kT) of the fireball from the best-fit model (\sw{CPL+BB}) along with the prompt emission light curve in the background. The panel (e) and (f) represent the evolution of magnetic field strength B and electron energy distribution index $p$ from the physical \sw{synchrotron} model.}
    \label{fig:prompt}
    \end{minipage}
\end{figure*}

\vspace{-0.6 cm}
Time-integrated spectrum analysis shows that the \sw{Band+BB} function provided the most accurate fit with the lowest DIC value. The obtained peak energy E${\rm pt}$ = 284$_{-5.56}^{+5.50}$, isotropic energy E${\rm \gamma,iso}$ = 4.31 $\times$ 10$^{53}$ erg, and peak luminosity L${\rm \gamma, iso}$ = 7.08 $\times$ 10$^{52}$ erg s$^{-1}$, for GRB 210610B are consistent with the well-studied Amati \citep{Amati} and Yonetoku \citep{Yonetoku_2004} correlations. The results of the time-resolved spectral analysis are shown in Figure \ref{fig:prompt}. A comparison of DIC values indicated that most of the bins are best fitted by the \sw{CPL+BB} model, while the physical \sw{synchrotron} model is favored by some of the bins. Spectral parameters obtained from the best-fit models in the time-resolved analysis show the following evolution pattern: The peak energy (E$_{\rm pt}$) and the magnetic field strength (B) are found to track the observed flux. Furthermore, the low energy spectral index ($\alpha_{\rm pt}$) is found crossing the synchrotron line of death and showing the hard to soft evolution, while the electron energy distribution index ($p$) has remained almost constant throughout the burst.\\

\begin{figure}
    \centering
    \includegraphics[scale=0.4]{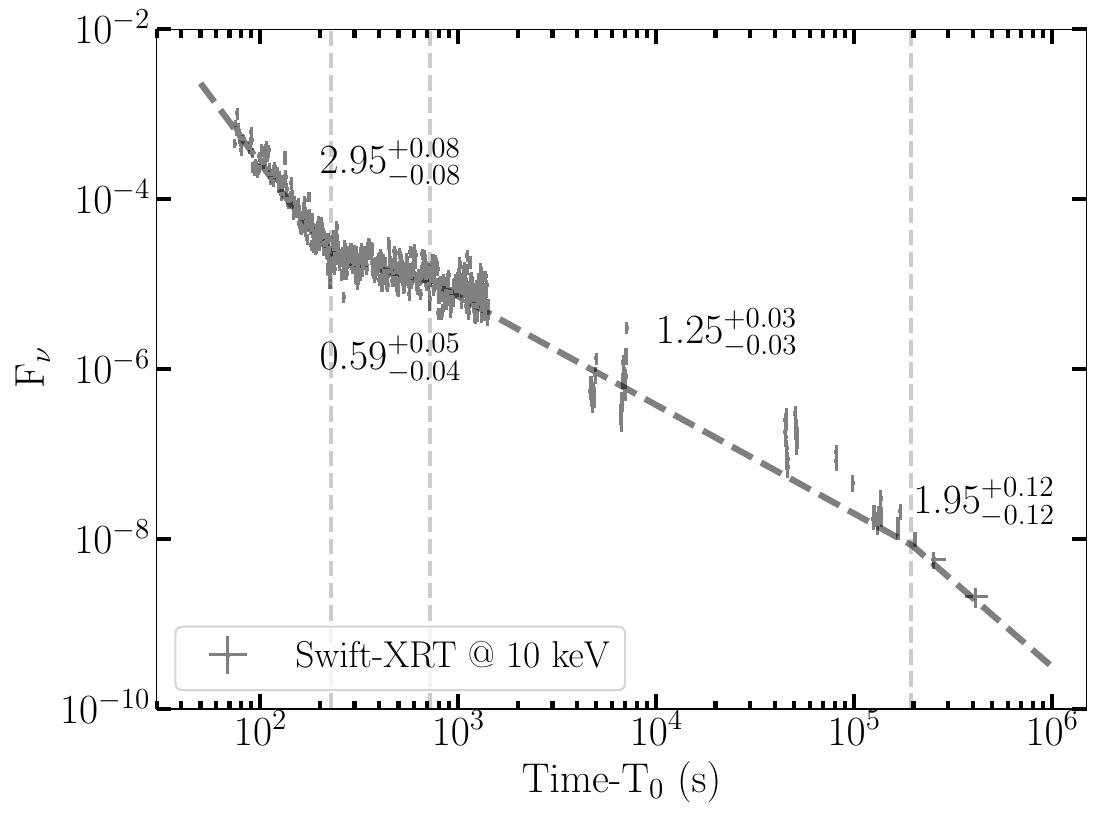}
    \includegraphics[scale=0.4]{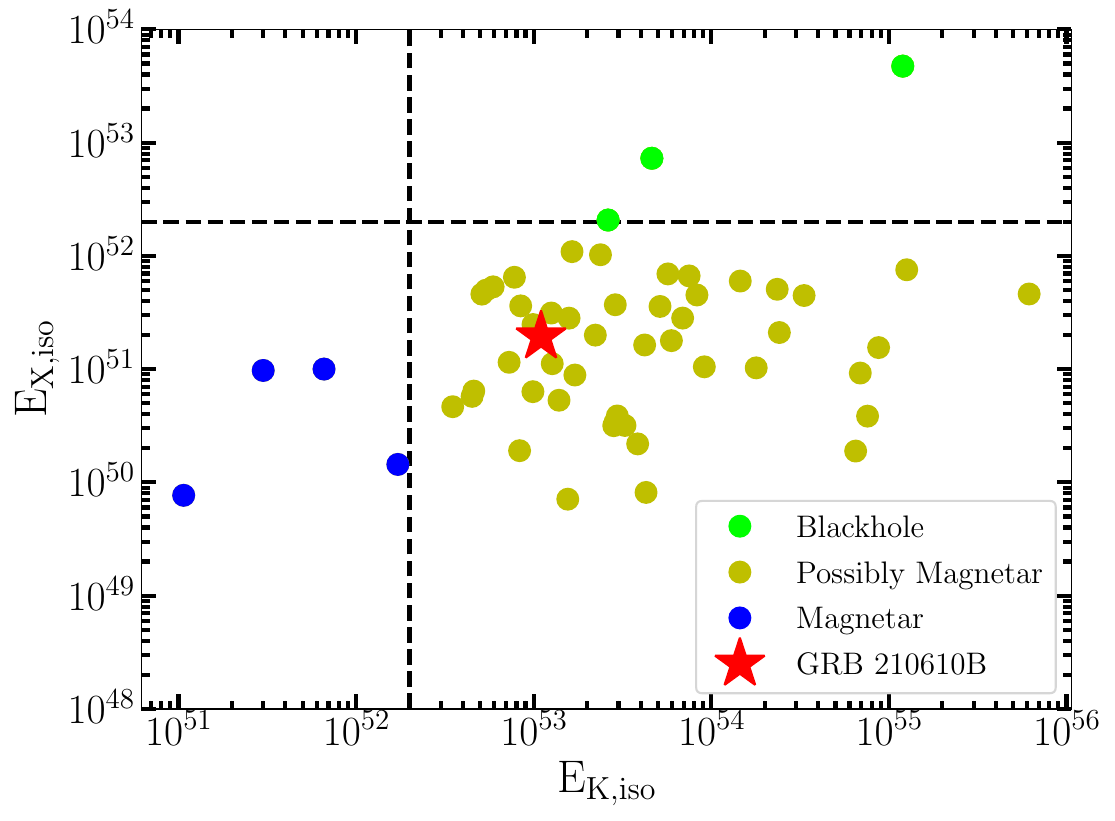}
    \bigskip
    \begin{minipage}{12cm}    
    \caption{Left: \swift-XRT flux density light curve of GRB 210610B. A 3-break power-law fitted to the light curve is shown with a dashed line. The decay indices are mentioned on their corresponding slopes. The three vertical dashed lines represent the time corresponding to the breaks in the light curve. Right: a comparison of observed energy released during the plateau phase (E$_{\rm X,iso}$) vs. total kinetic energy release (E$_{\rm K,iso}$) of the GRBs detected with \swift-XRT, taken from Ror et al., 2023 (in preparation).}
    \label{fig:oxlc}
    \end{minipage}
\end{figure}

\vspace{-0.5 cm}
\textbf{{\large X-ray Afterglow:}} To study the X-ray afterglow, we downloaded the \swift-XRT observations \citep{2021GCN.30170....1P, 2021GCN.30208....1G} from archive \sw{www.swift.ac.uk}. To fit the \swift-XRT light curve, we employed broken power-law models with one, two, and three breaks. The fitting was performed using the \sw{QDP} package, and the $\chi^2$ statistic was utilized to determine the best-fit model. Among these models, a three break power-law provided the best fit to the XRT light curve, and the obtained parameters are $\alpha_{\rm x1}$ = 2.95$_{-0.08}^{+0.08}$, $\alpha_{\rm x2}$ = 0.59$_{-0.04}^{+0.05}$, $\alpha_{\rm x3}$ = 1.25$_{-0.03}^{+0.03}$, $\alpha_{\rm x4}$ = 1.95$_{-0.12}^{+0.12}$, t$_{\rm xb1}$ $\sim$ 220, t$_{\rm xb2}$ $\sim$ 700, and t$_{\rm xb3}$ $\sim$ 1.95 $\times$ 10$^{5}$. The \swift-XRT observation and the best-fit model curve are shown in Figure \ref{fig:oxlc}.

\vspace{-0.5 cm}
\section{Discussion and Conclusion}
\textbf{{\large Origin of prompt emission:}} Time-resolved spectral analysis results of GRB 210610B show that $\alpha_{\rm pt}$ is beyond the synchrotron line of death. However, \cite{2020NatAs...4..174B} suggested that spectra can be well modeled with a synchrotron model even if the low-energy spectral index exceeds the synchrotron line-of-death. Indeed, we found that some of the bins are well fit by the synchrotron model. However, some of the bins favor the presence of superimposed thermal components as well. We studied the evolution of spectral parameters and found a rare feature where E$_{\rm pt}$ and B both showed flux-tracking behavior throughout the prompt emission. The observed feature can be explained in terms of fireball cooling and expansion \citep{2021MNRAS.505.4086G, 2023ApJ...942...34R}. In the light of above, we suggest that GRB 210610B has a hybrid jet (Poynting flux outflow moving along with a hot fireball) composition which results in the synchrotron emission superimposed over a thermal component \citep{2015AdAst2015E..22P}.\\

\vspace{-0.5 cm}
\textbf{{\large Progenitor:}}
GRB emission is highly collimated, therefore, the jet opening angle ($\theta_{\rm j}$) is a key parameter to get insights into its physics and energetics. One common method to estimate $\theta_{\rm j}$ is from the fitting of the observed afterglow light curve. The time corresponding to the sudden fall in the X-ray light curve from the normal decay phase is considered as the jet break time T$_{\rm j}$. Jet opening angle can be calculated using the relation: $\rm \theta_{\rm j} \sim 0.057 \times {\rm T_{\rm j, days}}^{3/8} \times (({1+z})/{2})^{3/8} \times ({E_{\gamma, iso,53}})^{1/8} \times (\epsilon_{0.2})^{1/8} \times ({n_{0.1}})^{1/8}$. GRBs with a T$_{90}$ duration longer than 2s are typically associated with the collapse of massive stars known as "collapsars". However, recent discoveries by \cite{2021NatAs...5..917A, 2022Natur.612..228T} have challenged this conventional understanding of the relationship between T$_{90}$ duration and GRB progenitors. To confirm whether the origin of GRB 210610B was indeed a collapsing massive star, we employed the relation presented in \cite{2011ApJ...739L..55B}: $\rm T_{\rm Bore} (s) \sim 15 \times \epsilon_{\gamma}^{1/3} \times ({L_{\rm \gamma,iso,50}})^{1/3} \times ({\theta}_{10^{\circ}})^{2/3} \times ({R_{11}})^{2/3} \times ({M}_{15M_{\odot}})^{1/3}$. Here, T$_{\rm Bore}$ represents the time required for the ultra-relativistic jet to penetrate the pre-existing cocoon surrounding the progenitor star, a method for calculation of T$_{\rm Bore}$ given in \cite{2023ApJ...942...34R}. T$_{90}$/T$_{\rm Bore}$ $>$ 1 suggests that the burst originated from a collapsing massive star. For GRB 210610B, T$_{90}$/T$_{\rm Bore}$ $\sim$ 107, providing strong evidence that the most probable progenitor of this burst was indeed a collapsing massive star.\\

\vspace{-0.5 cm}
\textbf{{\large Central engine:}}
One of the possible mechanisms for producing ultrarelativistic jets is through the formation of a highly rotating neutron star associated with extremely strong magnetic field lines (a millisecond magnetar). Pieces of evidence for a magnetar central engine are the existence of a plateau in the afterglow light curve and a highly polarized gamma-ray emission, associated with some GBRs, which could be produced by the strong magnetic field of a magnetar \citep{2001ApJ...552L..35Z}. To constrain the central engine responsible for GRB 210610B, we employed two methods.\\
\textbf{First method:} It involved the calculation of beaming corrected energy release during the prompt emission. To obtain this, we multiplied a beaming correction factor (f$_{\rm b}$) = 1 - $\cos({\theta_{\rm j}})$ $\sim$ 1/2($\theta_{\rm j})^{2}$ to the isotropic energy E$_{\rm \gamma,iso}$ of the burst i.e E$_{\gamma,\theta_{\rm j}}$ = f$_{\rm b}$ $\times$ E${\rm \gamma, iso}$. If the beaming corrected energy is greater than the maximum energy budget of a magnetar (i.e. E$_{\gamma, \theta_{\rm j}}$ $>$ 2 $\times$ 10$^{52}$), then it ruled out the possibility of a magnetar central engine \citep{2021ApJ...908L...2S}. However, for GRB 210610B, the beaming-corrected energy is E$_{\gamma, \theta_{\rm j}}$ = 1.06 $\times$ 10$^{51}$ erg $<$ 2 $\times$ 10$^{52}$ erg, suggesting that a magnetar could be the possible progenitor for this burst.\\
\textbf{Second method:} Under the assumption of synchrotron emission, the afterglow light curve of GRBs is expected to decay smoothly with a decay index of $\sim$ 1. However, some GRBs show a plateau in the afterglow light curve, which indicates that energy must be continuously supplied to the fireball to sustain the constant emission. A magnetar central is capable of providing such an energy injection \citep{2001ApJ...552L..35Z}. However, several other possible scenarios can explain the observed plateau phase. In some cases, a moderately relativistic classical fireball is enough to explain the plateaus in the X-ray light curve caused by an external shock in a low-density wind-like surrounding medium \citep{2022NatCo..13.5611D}. Further, GRB emission beamed narrowly in the forward direction with an opening angle $\theta_{\rm j}$ $\sim$ 1/$\Gamma$, with time, the Lorentz ($\Gamma$) factor of the jet decreases, and emission starts contributing from the off-axis region. This high-latitude emission can result in a plateau in the XRT light curve \citep{2020MNRAS.492.2847B, 2020ApJ...893...88O}. Furthermore, \cite{2008MNRAS.388.1729K} suggest that the continued accretion due to small viscous parameters and fall-back of residual gas on the central engine can also cause the plateaus in the X-ray light curve.\\
We have calculated the X-ray isotropic energy release, E$_{\rm X,iso}$, using the relation given in \cite{2018ApJS..236...26L}. A comparison between the energy released during the plateau phase (E$_{\rm X,iso}$) and the total kinetic energy release (E$_{\rm K,iso}$) of the GRBs detected by the \swift-XRT instrument is shown in Figure \ref{fig:oxlc} taken from Ror et al., 2023 (in preparation). For GRB 210610B, the obtained value of E$_{\rm X,iso}$ = 1.87 $\times$ 10$^{51}$ erg, which is less than the total energy budget of a magnetar, once again favoring the magnetar central engine as the likely progenitor for this burst. The combined energy emitted during the prompt+afterglow emission phase is 2.94 $\times$ 10$^{51}$ erg $<$ 2 $\times$ 10$^{52}$ erg favoring the magnetar central engine.\\

\vspace{-0.5 cm}
\textbf{{\large Future prospect:}} This article examines the characteristics of GRB 210610B through the utilization of archival data from space-based observations. Our forthcoming objective involves extending the analysis into the realm of multi-wavelength observations. Additionally, we intend to compare the afterglow emission of GRB 210610B with a collection of light curves from similar bursts to gain deeper insights into the burst's underlying progenitors and central engine.

\newpage

\begin{acknowledgments}
We thank the anonymous referee for his/her constructive and valuable comments. RG and SBP acknowledge the financial support of ISRO under AstroSat archival Data utilization program (DS$\_$2B-13013(2)/1/2021-Sec.2). AA acknowledges funds and assistance provided by the Council of Scientific \& Industrial Research (CSIR), India with file no. 09/948(0003)/2020-EMR-I. This research has used data obtained through the HEASARC Online Service, provided by the NASA-GSFC, in support of NASA High Energy Astrophysics Programs. AKR, SBP, RG, and AA also acknowledge the Belgo-Indian Network for Astronomy and Astrophysics (BINA) project under which local support was provided during the $3^{rd}$ BINA workshop.
\end{acknowledgments}
\\
\\
\\

\textbf{Appendix:}
\input{table1}
\input{table2}

\begin{furtherinformation}

\begin{orcids}
\orcid{0000-0003-3164-8056}{Amit}{Kumar Ror}
\orcid{0000-0003-4905-7801}{Rahul}{Gupta}
\orcid{0000-0002-9928-0369}{Amar}{Aryan}
\end{orcids}

\begin{authorcontributions}
Rahul Gupta has contributed to the prompt and afterglow analyses of the burst. S. B. Pandey and AA thoroughly reviewed the manuscript and provided insightful comments that greatly improved the quality of the draft.
\end{authorcontributions}

\begin{conflictsofinterest}
The authors affirm that they do not have any conflicts of interest regarding the manuscript and ensure that their research and findings are not influenced by personal or financial professional.
\end{conflictsofinterest}
\end{furtherinformation}

\bibliographystyle{bullsrsl-en}
\bibliography{extra}

\end{document}

%% file: table1.tex
\begin{table*}[ht!]
\footnotesize
\centering
\begin{minipage}{150mm}
\caption{Results from the prompt emission spectral fitting of GRB 210610B with \sw{CPL+BB} and \sw{Synchrotron} models.}
\vspace{0.2cm}
\label{tab:Prompt}
\end{minipage}
\begin{tabular}{|c|c||c|c|c|c||c|c|c|}\hline
\multicolumn{2}{|c||}{Time (s)} & \multicolumn{4}{c||}{CPL+BB} & \multicolumn{3}{c|}{Synchrotron}\\ \hline
T$_{start}$ & T$_{end}$ & $\alpha_{\rm pt}$ & E$_{\rm pt}$ (KeV)& KT (KeV)& DIC & B (G) & $p$ & DIC \\ \hline
11.94&19.43& {0.02}$_{-0.13}^{+0.13}$ & {256.27}$_{-16.41}^{+16.24}$ & {10.54}$_{-6.74}^{+6.60}$ &4524.82& {53.19}$_{-10.45}^{+10.6}$ & {3.69}$_{-0.25}^{+0.24}$ & 4605.47\\ 
19.43&22.19& {-0.18}$_{-0.10}^{+0.10}$ & {297.47}$_{-19.81}^{+20.01}$ & {10.11}$_{-6.35}^{+6.28}$ &3117.18& {55.05}$_{-9.44}^{+9.38}$ & {3.72}$_{-0.22}^{+0.21}$ & 3147.5\\ 
22.19&25.12& {-0.22}$_{-0.07}^{+0.06}$ & {328.98}$_{-14.35}^{+14.47}$ & {9.63}$_{-5.93}^{+6.20}$ &3324.12& {59.92}$_{-7.73}^{+7.29}$ & {3.88}$_{-0.10}^{+0.10}$ & 3410.32\\ 
25.12&26.51& {-0.20}$_{-0.06}^{+0.06}$ & {374.68}$_{-15.66}^{+15.61}$ & {10.10}$_{-6.32}^{+6.30}$ &2383.36& {70.27}$_{-8.73}^{+8.24}$ & {3.87}$_{-0.11}^{+0.10}$ & 2482.44\\ 
26.51&27.88& {-0.26}$_{-0.07}^{+0.07}$ & {451.70}$_{-20.28}^{+19.61}$ & {9.47}$_{-4.77}^{+4.98}$ &2433.52& {87.91}$_{-7.81}^{+7.8}$ & {3.95}$_{-0.05}^{+0.04}$ & 2658.72\\ 
27.88&29.94& {-0.28}$_{-0.05}^{+0.05}$ & {315.75}$_{-10.90}^{+11.03}$ & {10.08}$_{-6.33}^{+6.22}$ &2930.49& {55.39}$_{-5.18}^{+5.21}$ & {3.94}$_{-0.05}^{+0.05}$ & 2965.94\\ 
29.94&31.07& {-0.39}$_{-0.04}^{+0.04}$ & {478.23}$_{-18.66}^{+18.86}$ & {9.95}$_{-6.16}^{+6.19}$ &2120.92& {87.07}$_{-8.97}^{+9.32}$ & {3.94}$_{-0.05}^{+0.05}$ & 2239.31\\ 
31.07&34.3& {-0.41}$_{-0.03}^{+0.03}$ & {350.54}$_{-9.97}^{+10.06}$ & {9.70}$_{-6.07}^{+6.26}$ &3661.98& {56.78}$_{-4.28}^{+4.34}$ & {3.96}$_{-0.03}^{+0.03}$ & 3879.31\\ 
34.3&38.93& {-0.43}$_{-0.04}^{+0.03}$ & {290.16}$_{-9.00}^{+9.04}$ & {10.52}$_{-6.33}^{+6.08}$ &4032.09& {43.83}$_{-3.14}^{+3.39}$ & {3.95}$_{-0.04}^{+0.04}$ & 4114.57\\ 
38.93&41.69& {-0.47}$_{-0.05}^{+0.05}$ & {259.63}$_{-11.98}^{+12.07}$ & {9.99}$_{-6.18}^{+6.30}$ &3223.37& {36.23}$_{-5.25}^{+4.71}$ & {3.91}$_{-0.08}^{+0.08}$ & 2843.26\\ 
41.69&45.74& {-0.43}$_{-0.09}^{+0.11}$ & {316.10}$_{-21.33}^{+19.40}$ & {7.99}$_{-3.42}^{+4.35}$ &3776.79& {42.41}$_{-4.08}^{+4.48}$ & {3.93}$_{-0.06}^{+0.06}$ & 3199.64\\ 
45.74&47.41& {-0.55}$_{-0.06}^{+0.06}$ & {230.60}$_{-14.75}^{+14.68}$ & {10.51}$_{-6.59}^{+6.45}$ &2477.69& {29.21}$_{-4.33}^{+4.48}$ & {3.84}$_{-0.13}^{+0.13}$ & 2403.65\\ 
47.41&51.62& {-0.59}$_{-0.26}^{+0.15}$ & {174.65}$_{-31.82}^{+45.22}$ & {14.52}$_{-9.58}^{+10.31}$ &3458.08& {18.42}$_{-2.47}^{+2.3}$ & {3.92}$_{-0.06}^{+0.06}$ & 3638.75\\ 
51.62&60.63& {-0.60}$_{-0.06}^{+0.07}$ & {110.32}$_{-5.98}^{+5.81}$ & {10.33}$_{-6.46}^{+6.47}$ &4842.19& {13.12}$_{-1.52}^{+1.53}$ & {3.92}$_{-0.06}^{+0.06}$ & 4897.28\\ 
60.63&77.66& {-0.57}$_{-0.08}^{+0.08}$ & {93.04}$_{-5.35}^{+5.06}$ & {10.34}$_{-6.62}^{+6.81}$ &5598.38& {11.28}$_{-1.26}^{+1.31}$ & {3.93}$_{-0.06}^{+0.06}$ & 5697.72\\ 
77.66&92.29& {-0.59}$_{-0.31}^{+0.27}$ & {90.88}$_{-24.93}^{+15.82}$ & {11.63}$_{-5.36}^{+4.11}$ &5063.91& {8.123}$_{-1.44}^{+1.37}$ & {3.85}$_{-0.12}^{+0.12}$ & 5495.58 \\ \hline
\end{tabular}
\end{table*}

%% file: table2.tex
\begin{table*}[hbt!]
\footnotesize
\centering
\begin{minipage}{150mm}
\caption{Results obtained from various power-law (PL) fitted to the \swift-XRT lightcurve of GRB 210610B.}
\vspace{0.2cm}
\label{tab:afterglow}
\end{minipage}
\begin{tabular}{|p{0.8cm}|c|c|c|c|c|c|c|c|c|}\hline
 Model& $\alpha_{\rm x1}$ & $\alpha_{\rm x2}$ & $\alpha_{\rm x3}$ & $\alpha_{\rm x4}$ & t$_{\rm xb1}$ (s)& t$_{\rm xb2}$ (s)& t$_{\rm xb3}$ (s)& $\chi_{\nu}^{2}$ \\ \hline
PL & 1.14$_{-0.01}^{+0.01}$ &  &  &  &  &  &  & 8.81 \\ 
PL1 & 2.95$_{-0.06}^{+0.06}$ & 1.10$_{-0.01}^{+0.01}$ &  &  & 211$\pm$4.17 &  &  & 5.84 \\ 
PL2 & 3.01$_{-0.07}^{+0.07}$ & 0.54$_{-0.07}^{+0.07}$ & 1.29$_{-0.03}^{+0.03}$ &  & 231$\pm$5.27 & 715$\pm$51.48 & & 4.38 \\ 
PL3 & 2.95$_{-0.08}^{+0.08}$ & 0.59$_{-0.06}^{+0.06}$ & 1.25$_{-0.03}^{+0.03}$ & 1.95$_{-0.12}^{+0.12}$ & 220$\pm$6.22 & 710$\pm$55.11 & (1.96$\pm$1.52) $\times$ 10$^{5}$ & 4.36 \\ \hline
\end{tabular}
\end{table*}